# Diamond anvil cell using boron-doped diamond electrodes covered with undoped diamond insulating layer

Ryo Matsumoto[1,2]*, Aichi Yamashita[1,2], Hiroshi Hara[1,2], Tetsuo Irifune[3], Shintaro Adachi[1], Hiroyuki Takeya[1], and Yoshihiko Takano[1,2]

[1]*WPI-MANA, National Institute for Materials Science, Tsukuba, Ibaraki 305-0047, Japan*
[2]*University of Tsukuba, Tsukuba, Ibaraki 305-8577, Japan*
[3]*Geodynamics Research Center, Ehime University, Matsuyama, Ehime 790-8577, Japan*

E-mail: MATSUMOTO.Ryo@nims.go.jp

Diamond anvil cell using boron-doped metallic diamond electrodes covered with undoped diamond insulating layer have been developed for electrical transport measurements under high pressure. These designed diamonds were grown on a bottom diamond anvil via a nanofabrication process combining microwave plasma-assisted chemical vapor deposition and electron beam lithography. The resistance measurements of high quality FeSe superconducting single crystal under high pressure were successfully demonstrated by just putting the sample and gasket on the bottom diamond anvil directly. The superconducting transition temperature of FeSe single crystal was enhanced up to 43 K by applying uniaxial-like pressure.

High pressure is a promising tool to obtain new functional materials which cannot appear under ambient pressure, such as metallic hydrogen. Since hydrogen has the lightest atomic mass, the metallic hydrogen or hydrogen-rich compounds have been expected to be a candidate of high-transition temperature ($T_c$) superconductors for a long time [1-4]. Although the metallization of compressed hydrogen under 495 GPa was recently observed by optical analysis [5], the electrical transport measurements have never been conducted. On the other hand, the discovery with great surprise of high-$T_c$ superconductivity in hydrogen-rich material $H_3S$ at 203 K was reported under 150 GPa [6]. Many related studies have been continued after the discovery, for example, the crystal structural analysis [7], substitution effects [8,9], theoretical studies [4]. Especially, extremely high $T_c$ around 280 K is theoretically estimated under 250 GPa in sulfur-doped $H_3S$ [9].

A diamond anvil cell (DAC) is unique tool to generate a static high pressure above 100 GPa. If the electrodes are inserted in the sample space, the in-situ electrical transport measurement can be performed under high pressure. However, such a measurement under high pressure using DAC is quite difficult. In the case of DAC assembly for the $H_3S$ compression, the size of sample space was 10–30 μm [6], and four electrodes should be inserted into the narrow sample space to measure resistivity. Moreover, there is a risk in the DAC compression, which is deformation of the inserted electrodes. The development of an innovative technique is required for the measurements under high pressure.

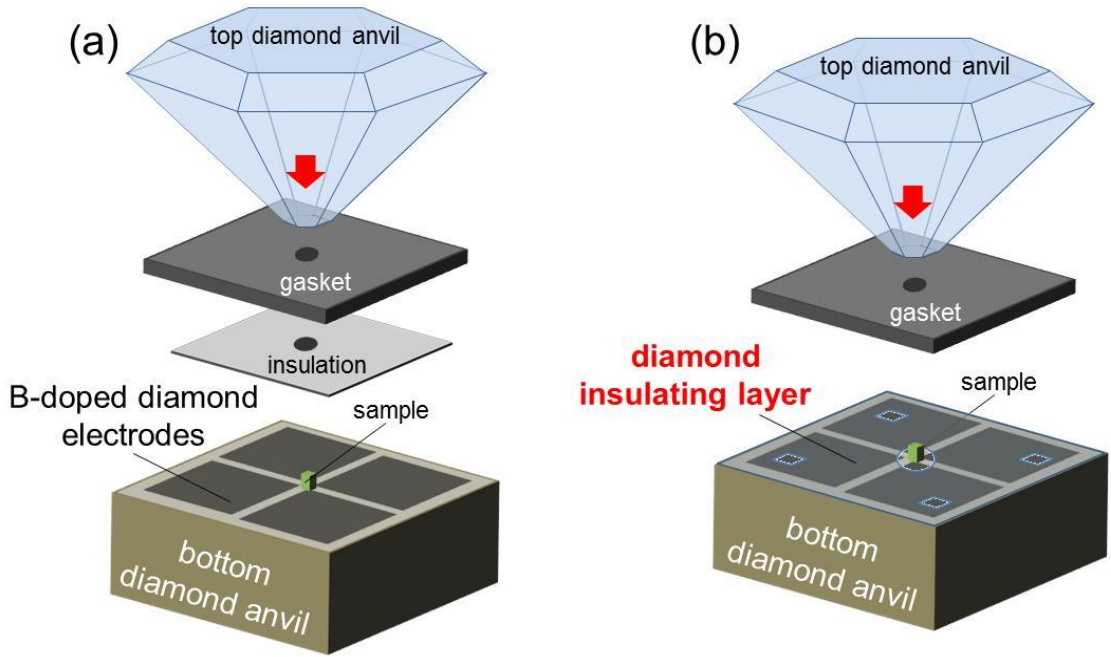

**Fig. 1.** (Color online) (a) Schematic of DAC assembly with boron-doped metallic diamond electrode for the electrical transport measurement. (b) Schematic of DAC assembly with undoped diamond insulating layer.

As shown in the Fig. 1(a), we have developed a DAC with boron-doped metallic diamond (BDD) electrodes on the bottom diamond anvil surface for the resistivity measurements under high pressure [10,11]. When the boron concentration in diamond exceeds $3\times10^{20}$ cm$^{-3}$, it shows metallicity and superconductivity at low temperature [12-16], keeping its high Vickers hardness [17]. The BDD electrodes were nanofabricated on the bottom anvil by combining a microwave-assisted plasma chemical vapor deposition (MPCVD) method and an electron beam lithography (EBL) technique [11]. The resistivity under high pressure can be easily measured by just putting sample on the BDD electrodes of the bottom diamond anvil. Moreover, the BDD electrodes can be used repeatedly until the anvil itself is broken [10].

However, there are not only the electrodes but also an insulating layer as a key component of DAC for electrical measurement under high pressure, to separate the electrodes from a metal gasket, as shown Fig. 1(a). The insulation is sometimes broken by applying pressure, even if hard materials are used, e.g., BN, $Al_2O_3$, and $CaSO_4$. It is necessary to develop harder insulating layer to measure resistivity under high pressure more easily. In this study, we focused on an undoped diamond (UDD) as a good candidate of the insulating material, which has a wide band gap of 5.4 eV and great Vickers hardness. We have fabricated the UDD insulating layer on the bottom diamond anvil using MPCVD and EBL. The electrical transport measurements of Fe-based superconductor FeSe single crystals were demonstrated by using developed DAC.

Figure 1(b) illustrates the schematic of DAC assembly with the BDD electrodes covered with the UDD insulating layer. The DAC is composed of only 3 parts; top anvil, bottom anvil, and metal gasket because the electrodes and insulating layer have been already drawn on the bottom diamond anvil. The electrical transport measurement can be more easily performed by just putting a sample and gasket on the bottom anvil directly. Moreover, we can use repeatedly not only the electrodes but also the insulating layer. The fabrication processes of UDD insulating layer on the BDD electrodes are described along the Fig. 2(a-f). (a)The BDD electrodes were prepared on commercially available substrate of IIa-type single crystal CVD diamond with (100)-crystal orientation by referring the previous reports [10,11]. The size of substrate was 2.5×2.5×1.2 mm. (b-c)The pattern of insulating layer was designed by a resist using EBL in combination with a scanning electron microscope (JEOL: JSM-5310) equipped with a nanofabrication system (Tokyo Technology: Beam Draw) on the BDD electrodes. (d-e)A Ti/Au metal mask was fabricated on the designed resist (Fig. 2(d)) through a lift-off process. The obtained metal mask was annealed at 450 ºC for 1 h in a furnace with Ar flow to make a TiC adhesion layer in the interface between the diamond substrate and the metal mask. (f)The homoepitaxial UDD was selectively grown on the uncovered region of the anvil surface by MPCVD using $CH_4$ diluted with $H_2$. The total pressure, total gas flow rate, and microwave power during the growth were maintained at 35 Torr, 400 sccm, and 350 W, respectively. The growth condition was 0.1% $CH_4/H_2$. The UDD thin film of 200 nm thick was grown from the anvil surface. (g)The UDD insulating layer was obtained after a wet etching treatment of Ti/Au metal mask using $H_2O_2$ and $NH_3$ mixture solvent.

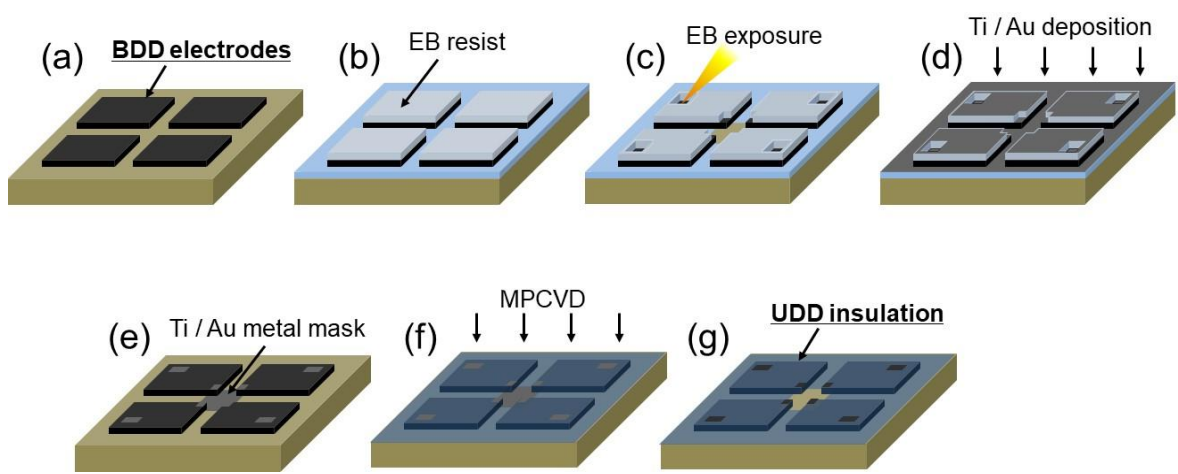

**Fig. 2.** (Color online) Nanofabrication process of the undoped diamond insulating layer. (a) Diamond substrate with boron-doped diamond electrodes. (b) Electron beam resist on the diamond substrate. (c) Design of the electrodes using EBL. (d) Deposition of the Ti/Au. (e) Metal mask after the lift-off process. (f) Deposition of the undoped diamond. (g) The undoped diamond insulating layer on the boron-doped diamond electrodes can be obtained after mask removal.

Figure 3(a) shows the optical microscope image of the obtained diamond anvil with BDD electrodes and UDD insulating layer. The BDD electrodes are covered by UDD thin film except for the circle area as a sample space in the center part and square areas as a terminal for current leads in the corner part of the anvil. The metal gasket can be put on the bottom anvil directly without additional intermediate layers because the UDD insulating layer separates the electrodes from the metal gasket. The electrodes are connected to the measurement systems such as current source and voltage meter from the square area using conductive paste and gold wire. Figure 3(b) shows the enlargement around the circle area as a sample space. A sample is put on the center of the circle, and then it is connected to the BDD electrodes. The diameter of the circle is 150 μm. The transport property such as the resistivity and the Hall coefficient can be measured by using five proves of electrodes. Moreover, a side-gating electric double-layer transistor can be formed on the diamond anvil using $V_g$ prove. If two proves of electrodes are used as a heater, high temperature synthesis of material, high temperature in-*situ* X-ray structural analysis, and high temperature Raman spectroscopy can be performed under high pressure. These functions are quite useful for not only the superconducting research but also wide fields of material science.

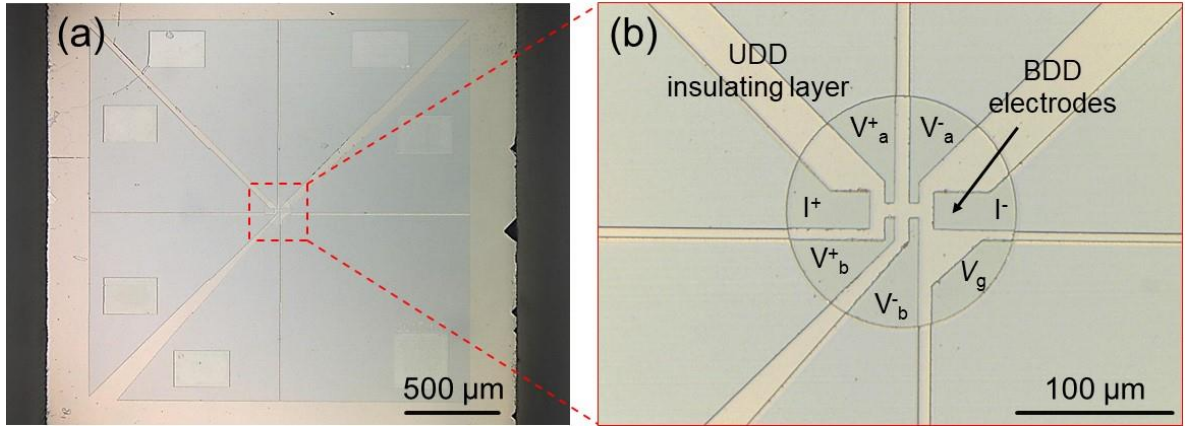

**Fig. 3.** (Color online) Diamond anvil with boron-doped diamond electrodes covered with undoped diamond insulating layer. (a) Entire image, (b) Enlargement of the sample space.

We demonstrate the electrical transport measurements of high-quality Fe-based superconductor FeSe single crystal under high pressure using the developed DAC. FeSe has the simplest crystal structure among the Fe-based superconductors [18] and it is expected to be a candidate material for superconducting wires and tapes [19, 20]. Although the original $T_c$ of FeSe is ~8 K, it is increased up to ~37 K by applying pressure [21-24] and ~ 46 K by using the intercalation of alkali or alkaline earth elements [25-29]. Moreover, extremely higher $T_c$ around 100 K was recently reported by using atomic layer FeSe thin film [30]. These suggest that the optimized FeSe is potential high-$T_c$ superconductor. Here, it is well-known that the $T_c$ of FeSe tends to enhance with decrease of Se height from Fe layer [22,31]. It is expected that the $T_c$ of single crystal FeSe could be dramatically enhanced by applying uniaxial pressure to compress the Se height effectively.

The high quality FeSe single crystal was prepared by using the chemical vapor transport method [32]. Fe and Se powders were mixed in an atomic ratio 1.1 : 1 and sealed in an evacuated quartz tube together with a eutectic mixture of KCl and AlCl$_3$. The ampoule was heated to 390 ºC on one end while the other end was kept at 240 ºC. After 28.5 days isometric

FeSe crystals with tetragonal morphology were extracted at the colder end. The obtained single crystals exhibited a plate-like shape that was 0.5-1.0 mm in size and 10-30 μm in thickness, and it showed $T_c$ around 8 K under ambient pressure via standard four-prove resistivity measurement.

The cleaved sample was put on the BDD electrodes as shown in the Fig. 4(a). The pre-pressed metal gasket of stainless steel sheet with a thickness of 200 μm was fixed on the bottom anvil as shown in the Fig. 4(b). Here, following (a–c) types of the DAC assembly were prepared to compare with hydrostatic and uniaxial pressure effects for FeSe. (a) A liquid pressure-transmitting medium of glycerin was used. In this assembly, the hydrostatic pressure could be induced in the sample space. (b) A solid pressure-transmitting medium of cBN was used to induce the uniaxial pressure to compress the Se heigh of FeSe single crystal. (c) cBN was used as both pressure-transmitting medium and insulating layer instead of UDD to emphasize the uniaxial component in the applied pressure. These anisotropic pressures would affect to the superconducting properties of the FeSe single crystal.

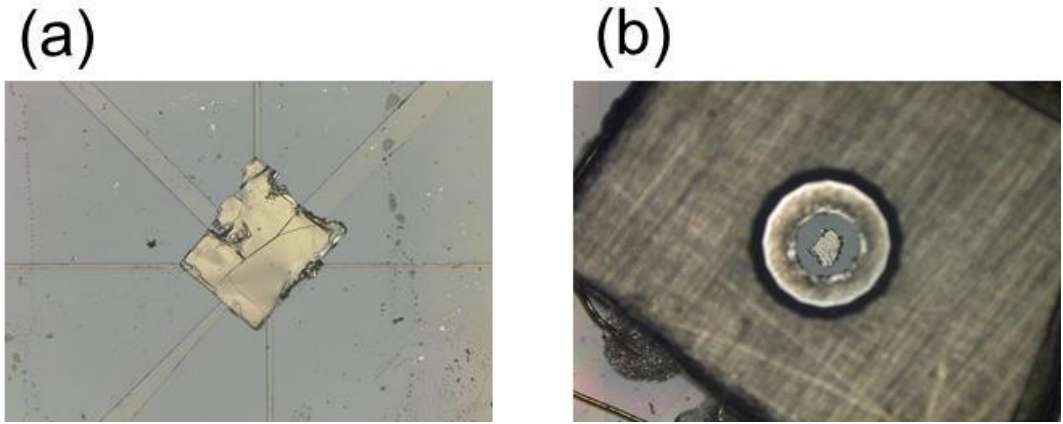

**Fig. 4.** (Color online) (a) FeSe single crystal on the boron-doped diamond electrodes. (b) Stainless-steel gasket on the bottom anvil.

The sample was compressed by squeezing the diamond anvil after the hole of gasket was filled by a pressure-transmitting medium with a manometer of ruby powder, and then the sample was electrically connected to the electrodes. The culet size of the top diamond anvil was 0.6 mm. The pressure values were determined from the peak position of the ruby fluorescence [33]. The fluorescence was detected by Raman spectroscopy system (RENISHAW: inVia Raman Microscope). The resistance of FeSe was measured by a standard four-terminal method using a physical property measurement system (Quantum Design: PPMS).

Figure 5 shows the temperature dependence of resistance in FeSe single crystal under various pressures using DAC assembly (a-c). In the compression (a), the onset temperature of superconducting transition ($T_c^{onset}$) was around 12 K under 0.4 GPa. It was drastically enhanced with increasing pressure, reaching its maximum value of 37 K at 10 GPa. The $T_c^{onset}$ was started to decrease from 11 GPa and completely disappeared at 18 GPa. This suppression of superconductivity was caused by a structural phase transition from an orthorhombic phase to a hexagonal phase [22]. The transition width was quite broad especially at high pressure regions. The broadening was possibly due to the pressure distribution in the sample between the voltage electrodes. The almost same behaviors were observed in the other compressions (b) and (c).

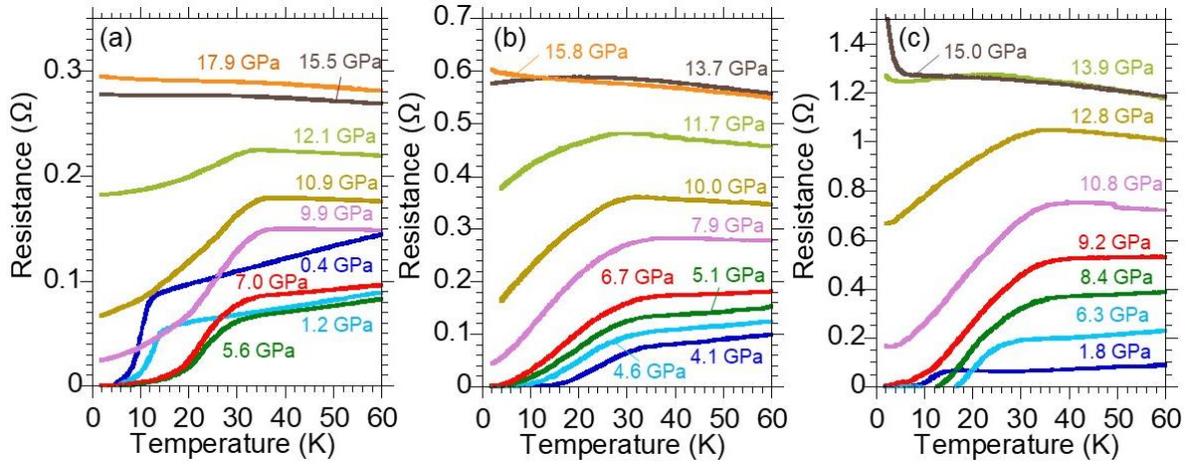

**Fig. 5.** (Color online) Temperature dependence of resistance in the FeSe single crystal under various pressures using DAC assembly (a-c).

Figure 6 shows the pressure dependences of $T_c^{onset}$ in the FeSe single crystal. The inset shows the temperature dependence of resistance around maximum $T_c^{onset}$. In the case of compression (a), the hydrostatic-like pressure would be induced in the sample space at least until 10 GPa which is freezing point of glycerin, because the sample space was completely sealed without any intermediate materials between the gasket and bottom anvil. The maximum $T_c^{onset}$ of FeSe was observed around 37 K, which was good agreement with previous report using polycrystal [22]. On the other hand, since the hard material cBN was used as the medium in the compression (b), it is suggested that the uniaxial-like pressure was generated to the sample with a direction against the Se height. The maximum $T_c^{onset}$ was changed, and then it situated around 39 K. It could be understood that the anion-height from Fe-layer of our single crystal sample was effectively decreased by applying uniaxial-like pressure. Finally, in the compression (c), it is expected that much emphasized uniaxial-like pressure was occurred in the sample space due to the deformable cBN insulating layer. The higher maximum $T_c^{onset}$ was remarkably situated around 43 K. These facts show that the $T_c$ of FeSe can be changed depending on the direction of applying pressure, and the developed DAC has selectivity of the hydrostatic and uniaxial component in the applied pressure.

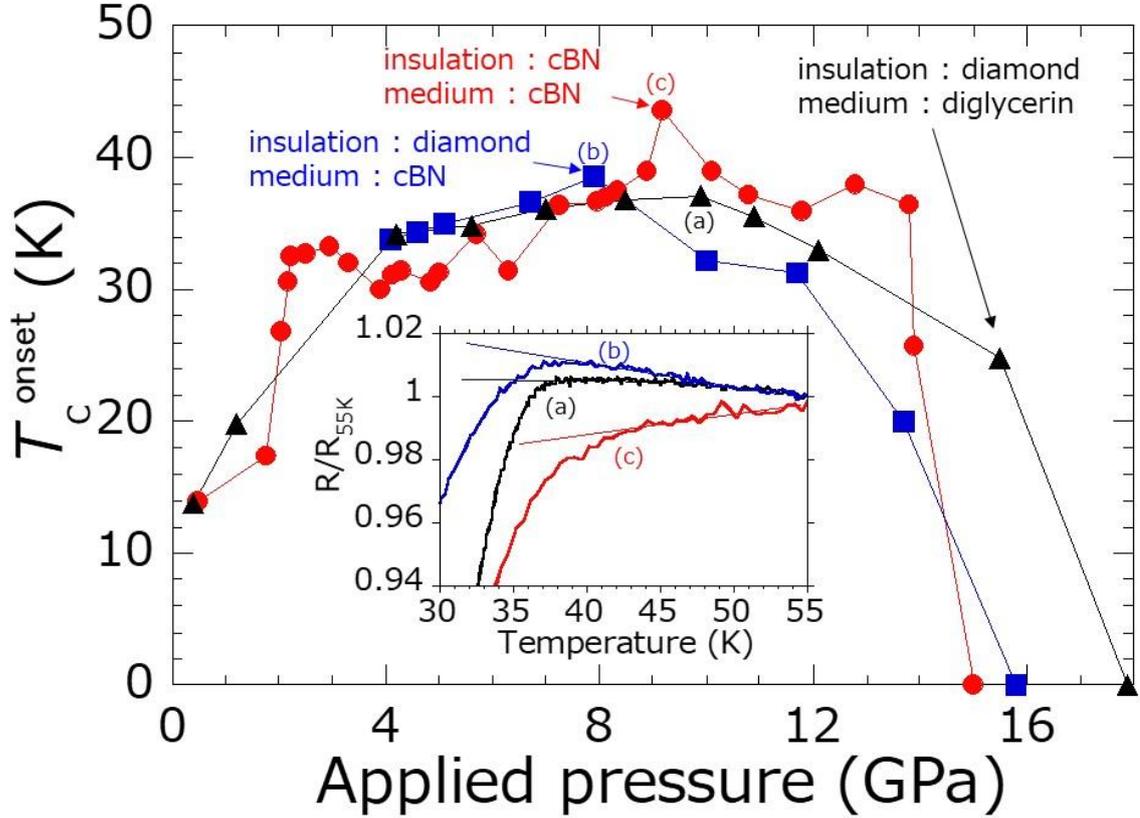

**Fig. 6.** (Color online) Pressure dependences of $T_c^{onset}$ in the FeSe single crystal using DAC assembly (a-c).

In summary, the DAC with BDD electrodes covered by UDD insulating layer has been developed by using a combination of MPCVD and EBL. The pressure dependence of $T_c^{onset}$ in high quality FeSe single crystal was investigated under hydrostatic and uniaxial-like pressures. Under the hydrostatic-like pressure, FeSe showed $T_c^{onset}$ around 37 K, which consistent with previous report using polycrystal sample. On the other hand, higher $T_c^{onset}$ was observed around 43 K under the uniaxial-like pressure with a direction of Se height of FeSe. It suggested that the anion-height from Fe-layer was effectively compressed. The developed DAC has selectivity of the hydrostatic and uniaxial component in the applied pressure.

**Acknowledgments**

This work was partly supported by JST CREST, Japan, and JSPS KAKENHI Grant Number JP17J05926. A part of the fabrication process was supported by NIMS Nanofabrication Platform in Nanotechnology Platform Project sponsored by the Ministry of Education, Culture, Sports, Science and Technology (MEXT), Japan. The part of the high pressure experiments were supported by the Visiting Researcher's Program of Geodynamics Research Center, Ehime University.